\documentclass[twocolumn,secnumarabic,amssymb,nobibnotes,showpacs]{revtex4}
\usepackage{amsmath}
\usepackage{latexsym}
\usepackage{float}
\usepackage{amssymb}
\usepackage{graphicx}
\usepackage{textcomp}
\usepackage{hyperref}
\def\ba{\begin{eqnarray}}
\def\ea{\end{eqnarray}}
\def\be{\begin{equation}}
\def\ee{\end{equation}}

\def\bm{\begin{math}}
\def\me{\end{math}}

\newcommand{\dummy}

\begin{document}

\title{Phase Behavior of Active Swimmers in Depletants: Molecular Dynamics and Integral Equation Theory}

\author{
          Subir K. Das$^{1}$, Sergei A. Egorov$^{2}$, Benjamin Trefz$^{3,4}$, Peter Virnau$^{3}$ and
          Kurt Binder$^{3}$}

\affiliation{$^{1}$ \textit{Theoretical Sciences Unit, Jawaharlal Nehru Centre for Advanced Scientific Research,
            Jakkur P.O., Bangalore 560064, India}\\
            $^{2}$ \textit{Department of Chemistry, University of Virginia, Charlottesville, Virginia 22901, USA}\\
            $^{3}$\textit{Institut f\"ur Physik, Johannes Gutenberg-Universit\"at, Staudinger Weg 7, 55128
                 Mainz, Germany}\\
	    $^{4}$\textit{Graduate School Material Science in Mainz, Staudinger Weg 9, 55128 Mainz, Germany}
}

\date{\today}

\begin{abstract}
We study the structure and phase behavior of a binary mixture where one of the
components is self-propelling in nature. The inter-particle interactions in the
system were taken from the Asakura-Oosawa model, for colloid-polymer mixtures,
for which the phase diagram is known. In the current model version the colloid
particles were made active using the Vicsek model for self-propelling particles. The
resultant active system was studied by molecular dynamics methods and
integral equation theory.
Both methods produce results consistent with each other and
demonstrate that the Vicsek model based activity facilitates phase separation, thus
broadening the coexistence region.
\end{abstract}
\maketitle

Various phenomena involving systems containing active particles have been of
significant recent research interest \cite{vic1,vic2,dre,how,pal1,zha,ke,cal,wan,loi,val,butt,bia,joa,sch,cat,kai,pal2,sun}.
Simple examples are flocking of birds \cite{vic1,vic2}, dynamics in a bacterial colony \cite{val}, etc.
Self-propelling character of active species make such systems extremely complex.
While the literature in this area gained significant volume in recent time,
many basic questions related to both equilibrium and nonequilibrium statistical
mechanics remain open. Examples \cite{onu} are phase behavior and criticality,
various fluctuation relations, kinetics of phase transitions, etc.

Considering the current status of the subject, realistic modeling involving shape,
flexibility, etc., of the constituents, is very difficult. Definition of temperature
in such systems is questionable and is a subject of much interest \cite{wan,loi}.
These systems are nonequilibrium
in nature and so, to what extent methodologies of equilibrium statistical mechanics
are applicable, e.g., to study phase behavior, remains to be seen. Thus, it is extremely
challenging to study active matter from
theoretical and computational points of view.

Interesting experiments \cite{sch,pal2,sun} reported that
addition of active particles in a system can dramatically alter its phase behavior.
With that objective, in
this letter, we present results for the phase behavior from molecular dynamics \cite{fre} (MD) and
integral equation theoretical \cite{han} (IET) study of a model where an active species
swims in an environment of passive particles.

Interactions among active particles can be of various types \cite{sch,pal2}.
In this work we are interested in constructing a model with an effective attraction among self-propelling
particles, which exhibits phase separation in the equilibrium limit. Thus, our objective
is to study the influence of dynamic clustering in active particles, on the
phase behavior of the corresponding passive system. Our results, in conjunction with Ref. \cite{sch},
demonstrate that, based on the type of activity, phase separation can either be facilitated or suppressed.
In the Vicsek type model, the tendency of active particles to form clusters enhances
phase separation.

In our model, the standard interactions
among particles are given via a variant of the well-known
Asakura-Oosawa (AO) model \cite{asa1,zau} for colloid ($A$) and
polymer ($B$) mixtures. In this work, we make the colloids self-propelling motivated
by the fact that bacteria or other active objects in solutions are in the size range
of colloidal particles in colloidal dispersion. Further, in many circumstances (e.g., living objects)
there are also depletants in the solution which may create attractions among
colloidal particles. It is then an intriguing question, what is the consequence
of the interplay between such ``ordinary'' interactions and the fact that one deals with
active, rather than passive particles. Moreover, colloid-polymer mixtures are popular for the study of phase
transitions in condensed matter because of the easy
observability of interfacial phenomena on the single particle scale. While the
original AO model \cite{asa1} cannot be straightforwardly generalized
(it is only defined in thermal equilibrium due to the ideal gas character of the
penetrable soft spheres representing the random walk-like polymers),
we use a variant \cite{zau} with nontrivial potentials.
Interparticle interactions in this model \cite{zau} are
described as ($r=|\vec{r_i}-\vec{r_j}|$, $\vec{r_i}, \vec{r_j}$ being the locations
of $i$th and $j$th particles)
\begin{equation}\label{pot1}
U_{\alpha \beta}=4\epsilon_{\alpha \beta}\left[\left(\frac{\sigma_{\alpha \beta}}{r}\right)^{12} -\left(\frac{\sigma_{\alpha \beta}}{r}\right)^{6} +\frac{1}{4}\right],
\end{equation}
where $\alpha \beta=AA$ or $AB$, while 
\begin{equation}\label{pot3}
\scriptstyle
U_{BB}=8\epsilon_{BB} \left[1-10\left(\frac{r}{r_{c,BB}}\right)^{3}+15\left(\frac{r}{r_{c,BB}}\right)^{4}-6\left(\frac{r}{r_{c,BB}}\right)^{5}\right],
\end{equation}
with $\epsilon_{AA}=\epsilon_{AB}=1$, $\epsilon_{BB}=0.0625$, $\sigma_{AA}=1$,
$\sigma_{AB}=0.9$, $\sigma_{BB}=0.8$ and $r_{c,\alpha\beta}$ $=2^{1/6}\sigma_{\alpha\beta}$.
Similar size ratios are conceivable in colloidal dispersions with biopolymers as depletants \cite{jam}.
Equilibrium phase behavior of this entirely passive model exhibiting entropy driven phase separation
was estimated with good accuracy \cite{zau}. As stated above,
in the present work, we introduce activity, via
the well-known Vicsek model \cite{vic1,vic2} as described below, to observe how the phase behavior of the AO model changes.

The Vicsek model was introduced to study self-propelled dynamics in interacting
biological systems where the interaction of a particle with its neighbors is
decided by the average direction of motion of the neighbors. We have used
$\sqrt{2}r_{c,AA}$ as a cut-off radius to define neighborhood. At each instant of
time, an active particle was supplied with an acceleration $f_A$, in addition to
the one due to interatomic interactions Eqs. (\ref{pot1},\ref{pot3}),
in the direction of the average velocity
of its neighbors. We study this model by MD 
as well as IET calculations.

When activity is introduced, eventually the temperature assumes a value higher than the one
initially assigned. This is due to additional velocity imparted according to Vicsek rule. A reason for
choosing the AO model as the passive limit is its weak sensitivity to temperature. In real systems the solvent carries
off all the heat produced by active particles. This effect, in our study has been avoided by an appropriate choice of the model.
Furthermore, in the AO model the polymer reservoir packing fraction, which regulates the depletant density, plays the role that inverse temperature would play in a standard system where competition
of energy and entropy controls the phase behavior.

In our MD simulations, we have implemented a Langevin thermostat to control the
overall temperature. There we have essentially solved the equation
\begin{equation}\label{langevin}
m\ddot{\vec{r}}_i=-\vec{\nabla} U -\gamma m \dot{\vec{r}}_i + \sqrt{2\gamma k_B T m} \vec R(t),
\end{equation}
particle mass $m$ being set to be same for all particles, $\gamma$
is a damping coefficient, $U$ is the interparticle potential, $\vec R$ is a $\delta$-correlated
noise and $T$ the temperature. Eq. (\ref{langevin}) was solved via implementation of
the velocity Verlet algorithm \cite{fre} with time step $\Delta t=0.002$ in units of
$\sqrt{\sigma_{AA}^2m/\epsilon_{AA}}$. Henceforth all lengths
are measured in units of $\sigma_{AA}$ and energy in units of $\epsilon_{AA}$. Further,
for convenience, we set $m$, $\sigma_{AA}$, $\epsilon_{AA}$ and Boltzmann's
constant $k_B$ to unity. To introduce activity we have used $f_A=10$, unless mentioned otherwise. Throughout the
work the starting temperature was $T=1$ with $\gamma=1$. For reasons mentioned
earlier, this temperature does not correspond to the effective temperature in the active case.
The thermostat has the effect to maintain a steady state with an enhanced temperature
of the colloids after a rather short transient. Thus no artificial temperature rescaling
is needed to compensate for continuous energy pumping due to the Vicsek model.
Also, we have checked using other values of $\gamma$, viz., $0.5$ and $2.0$, that our
results on the phase diagram are independent of these choices within
statistical errors.

An IET approach \cite{han} was adopted that
uses standard Ornstein-Zernike \cite{han} (OZ) equations relating direct pair correlations with the total
correlation functions. It is well-known that asymmetric models, as the one used here 
pose significant difficulty in the IET approach in obtaining closure relations
needed to supplement the OZ equations. To overcome this difficulty, we took
a pragmatic approach and empirically determined the best combination of closures by
comparing structural and thermodynamic output from the IET with the MD data at a
reference state. We found that the modified hypernetted chain \cite{han} closure for AA and the
Percus-Yevick \cite{han} closure for AB as well as BB work best and so results are presented
using these closures.

\begin{figure}[htb]
\centering
\includegraphics*[width=0.35\textwidth]{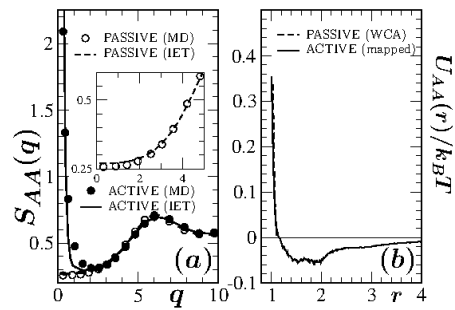}
\caption{
(a) Plots of the colloid-colloid structure factors for active and
passive systems for the state point $\eta_A=0.15$, $\eta_B=0.05$. Molecular
dynamics results are shown in symbols, IET
calculations are shown in lines. The inset shows a
magnified plot of passive system results.
(b)  Actual Weeks-Chandler-Anderson (WCA) colloid-colloid pair potential of the passive
system (solid line) and the corresponding effective potential of the
passive system (broken line) onto which the active one is mapped.
Note that $U_{AB}$ and $U_{BB}$ are the same for both the active and the passive case.
}\label{fig1}
\end{figure}

In order to apply IET methods to the active system,
we assume that it can be mapped onto
an effective passive one, interparticle interactions
chosen in such a way as to reproduce the structural properties of the
active system. If this assumption holds true, then both structure
and phase behavior of the active
system can be studied by applying the IET formalism outlined above to
the passive system, onto which the active system has been
mapped. In order to test our assumption, we selected a {\em single point}
on the phase plane [that corresponds to Fig.\ref{fig1}(a)]
and ``inverted'' the simulated radial
distribution function $g_{AA}$ of the active system  to
obtain an effective interaction potential for the passive system
mimicking the active one ($U_{AB}$ and $U_{BB}$ are the same as in the MD simulations).
For the passive case all interactions are given by Eqs.(\ref{pot1},\ref{pot3}).
We then performed  IET
calculations for the effective passive system over the entire phase plane.
The validity of our assumptions was confirmed by good agreement between
IET and MD results for the phase diagram of the active system.

\begin{figure}[htb]
\centering
\includegraphics*[width=0.4\textwidth]{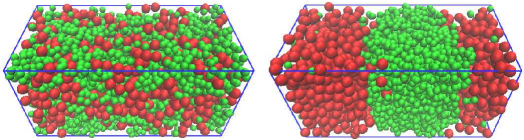}
\caption{
Snapshots from MD simulations for the state point $(\eta_A,\eta_B)$=(0.15,0.25).
Each rectangular box has linear dimension $L_z = 24$ ($L_x = L_y = 12$) in the elongated direction containing $945$ $A$ particles and $3078$
$B$ particles.
The left frame shows a representative equilibrium configuration for the passive system
whereas the right one is from the steady state of the active model. In both the pictures, the colloid
particles are shown in red and polymers in green.
}\label{fig2}
\end{figure}

\begin{figure}[htb]
\centering
\includegraphics*[width=0.35\textwidth]{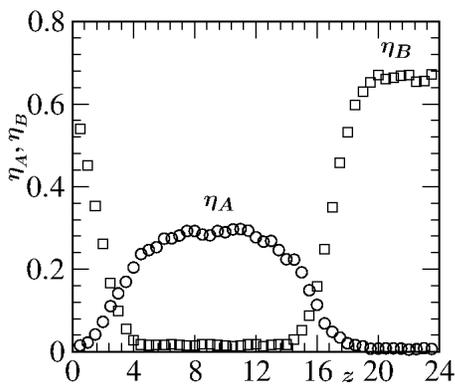}
\caption{
Profiles of $\eta_A$ and $\eta_B$, for an active system, along the elongated ($z$) direction
of a rectangular box containing $945$ $A$ and $3078$ $B$ particles.
This system phase separates into
$A$- rich and $B$- rich regions.
}\label{fig3}
\end{figure}

Via MD simulations and IET, in this letter we compare the results for structure and phase behavior
of the active model with that of the corresponding passive one \cite{zau}. All our MD
results were obtained using periodic boundary conditions starting from random initial
configurations. Both MD and IET results confirm that the miscibility gap of the active model widens.

In Fig. \ref{fig1} (a) we compare the structure factors of the active and passive models, calculated as
\begin{equation}\label{sofq}
S_{\alpha\beta}(q)=\frac{1}{N}\left<\sum\limits_{i=1}^{N_{\alpha}} \sum\limits_{j=1}^{N_{\beta}}
\mbox{exp}(i\vec{q}\cdot\vec{r}_{ij})\right>,
\end{equation}
for $\alpha=\beta=A$. In Eq. (\ref{sofq}),
$N_{\alpha}$, $N_{\beta}$ are respectively the number of particles of species $\alpha$, $\beta$ and
$N$ is the total number of particles. We estimate the
phase behavior by varying the composition of $A$ and $B$ particles. Previously \cite{zau},
for the passive model, the phase behavior was obtained in $\eta_A$ vs $\eta_B$ plane,
$\eta_{\alpha}$ being the packing fraction of species $\alpha$. To be more specific
$\eta_A=0.5484 \rho_A$ and $\eta_B=0.2808 \rho_B$,
where $\rho_{\alpha}$ ($=N_{\alpha}/V$, $V$ being the system volume) is the density of species
$\alpha$. The state point in Fig. \ref{fig1} (a) is ($\eta_A,\eta_B$) $\equiv$ ($0.15,0.05$).
For the passive case \cite{zau}, the critical point ($\eta_A,\eta_B$) is
at ($0.15,0.328$).
The symbols in this figure show the MD results and the IET calculations are shown
by lines. The agreement between MD and IET calculations must be appreciated.
Interestingly, the sharp rise of $S_{AA}$ at small $q$ for the active model indicates the
presence of long wavelength fluctuations not present in the passive system.

\begin{figure}[htb]
\centering
\includegraphics*[width=0.35\textwidth]{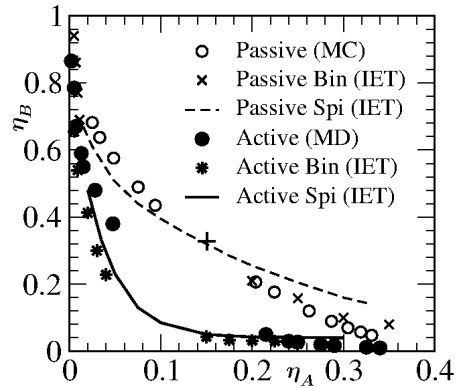}
\caption{
Phase behavior in the $\eta_A$-$\eta_B$ plane for both passive and active systems. Simulation and IET results
[spinodal (Spi) and binodal (Bin)] are shown. The simulation data
for the passive system are from previous \cite{zau} grandcanonical Monte Carlo (MC) work
whereas the ones for the active system are obtained from the MD method described in the text. The
plus (+) shows the critical point of the passive system.
}\label{fig4}
\end{figure}

Fig. \ref{fig1} (b) presents a comparison between the interaction potential
of the passive system and the effective interaction potential obtained via inversion of the
simulation data for the active system. Intriguingly, an additional attraction is present
in the active case, in contrast to the model of Ref. \cite{sch}. This provides an
explanation for the widening of the coexistence region seen below.

In Fig. \ref{fig2}, from MD simulations,  we show two snapshots from the passive and active models at a state point different from
Fig. \ref{fig1}(a) and closer to the passive coexistence curve (see Fig. \ref{fig4}).
The left snapshot shows the passive system, the right one is for the active model.
It can clearly be seen that, as opposed to the passive one, the active system has nicely phase separated.

\begin{figure}[htb]
\centering
\includegraphics*[width=0.4\textwidth]{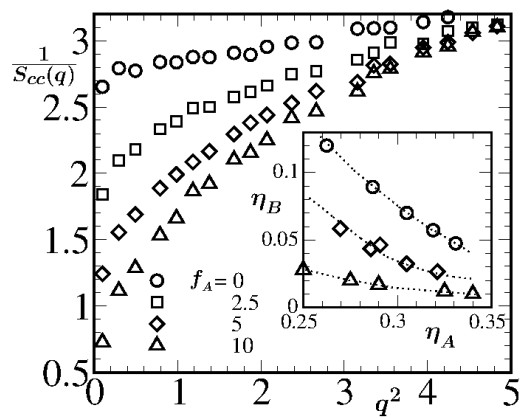}
\caption{
Ornstein-Zernike plot of concentration -concentration structure factor at the state point of Fig.\ref{fig1}. Results for few different
values of activity strength $f_A$ are included. The inset shows the phase diagram (compare with Fig.\ref{fig4}) for
three different values of $f_A$ in the colloid-rich region. The dotted lines are guides to the eye. $f_A = 5$ is located 
in between $f_A = 10$ and the passive case.
}\label{fig5}
\end{figure}

Next we estimate the phase diagram. To achieve this objective via MD simulations,
we have started with a homogeneous mixture of $A$ and $B$ particles
and waited for the system to reach a steady state which can be identified from
the potential energy and density profiles (see Fig~\ref{fig3}). If the snapshots showed phase separation in the
steady state, we have calculated profiles for $\eta_A$ and $\eta_B$ as a function of $z$.
Staying away from the interfacial
region, we extract the values of $\eta_A$ and $\eta_B$ for both coexisting phases \cite{wat}.
On the other hand, for the IET method, a divergence criterion for the structure factor in $q\rightarrow 0$
limit was used to obtain the spinodal.

Finally, our central result, the phase diagram, is presented in Fig. \ref{fig4}
in the $\eta_A$-$\eta_B$ plane. The result from the passive system \cite{zau}
is also added for comparison. It is clearly seen that the introduction of a Vicsek type
activity facilitates phase separation. Simulation and IET results are
in qualitative agreement for both the passive and the active case.
Some quantitative disagreement
that appears between the theory and simulation can possibly be
attributed to the mean field nature of the IET.
The IET result for the binodal was obtained following the procedure
outlined in Ref.~\cite{dzubiella}. As expected, it lies outside the
spinodal, approaching it in the critical region. Comparing the active and passive cases, it is
seen that the coexistence region opens up when activity is introduced. But strong finite-size effects and critical slowing down prevent us from obtaining points very close to the critical point reliably via MD simulations with moderate system
sizes.

It is worth noting that different types of activity can lead to phase behavior which is not
only quantitatively but also qualitatively different. In Ref. \cite{sch} activity
leads to randomly enhanced mobility resulting in an additional effective repulsion. In a Vicsek type
model, originally introduced to study swarming behavior, active particles tend to
cluster. This leads to an additional attractive interaction which enhances phase separation.

So far our results refer only to $f_A=10$.
In Fig. \ref{fig5} we show the inverse concentration-concentration structure factor $S_{cc}(q)$,
defined as ($x_{\alpha}=N_{\alpha}/N$)
\begin{equation}
S_{cc}(q)=x_B^2S_{AA}(q)+x_A^2S_{BB}(q)-2x_Ax_BS_{AB}(q),
\end{equation}
for different $f_A$. For small enough $q$ the OZ behavior \cite{han}
\begin{equation}
S_{cc}\approx k_BT\chi/(1+q^2\xi^2),
\end{equation}
with $\chi$ and $\xi$ being the susceptibility and correlation length, is nicely visible. The stronger
enhancement of $S_{cc}(q)$ with the increase of $f_A$ is suggestive of the fact that the phase gap widens
which is due to the stronger effective attraction among active particles \cite{angela}. This is directly demonstrated
in the inset.

In conclusion, we have presented results for the structure and phase behavior of a physically
motivated model mixture containing
active particles. The results are compared with the corresponding passive systems. Our molecular dynamics simulations and
integral equation theory (this being the first time in literature to have been applied for the
study of active matter) are in reasonable agreement and show that the
tendency to form clusters in the Vicsek model leads to an effective attraction
among colloids and enhances phase separation. This result is in contrast with previous work
in which the model of activity differs. Basically, depending upon how motility is introduced,
qualitatively different trends result. While it needs to be seen whether the 
conclusions hold true in more general situations, our results should be stimulating for future experiments.

It would be interesting to study critical phenomena for such phase separating active systems, and
to look at interfacial properties and hydrodynamic effects.
Analogous to the observation in phase separating passive fluids, our preliminary study of an active system also suggests fluctuation induced broadening of interfaces with increasing system size.
Despite
many interesting works involving active matter, to the best of our knowledge, understanding of such
important aspects is missing. We hope our work will influence experimentalists and theorists to
pursue such problems.

{\bf Acknowledgment:} SKD and SE acknowledge research visits to the Johannes Gutenberg-Universit\"at
in Mainz. SKD is also grateful for financial support from DFG (SPP1296, TR6/A5) and JGU, Germany, JNCASR
and DST, India, as well as ICTP, Italy. SKD and PV acknowledge discussion with F. Schmid. PV acknowledges discussions with M. Oettel.
P.V. and B.T. would also like to acknowledge the MAINZ Graduate School of Excellence.

\end{document}